\documentclass[aps,prd,superscriptaddress,showpacs,twocolumn,floatfix]{revtex4}

\usepackage{graphicx}% Include figure files
\newcommand{\beq}  {\begin{equation}}
\newcommand{\eeq}  {\end{equation}}
\newcommand{\bmath}{\begin{eqnarray}}
\newcommand{\emath}{\end{eqnarray}}

\def\lapproxeq{\lower .7ex\hbox{$\;\stackrel{\textstyle<}{\sim}\;$}}
\def\gapproxeq{\lower .7ex\hbox{$\;\stackrel{\textstyle>}{\sim}\;$}}
\def\be{\begin{equation}}
\def\ee{\end{equation}}
\begin{document}
%%%% INFORMATION FOR THE TITLE PAGE GOES IN HERE:%%%%%%%%%%%%%%%%%%%%%%%%%%%%%

\title{
Spin asymmetries for events with high $p_T$ hadrons in DIS
and an evaluation of the gluon polarization

\vspace{1.0cm}
\normalsize

{\it  In remembrance of Vernon W. Hughes, \\
initiator of the SMC experiment and spokesman of the collaboration, \\
who passed away on March 25, 2003  \\
and to whom this article is dedicated.\\}

\vspace{0.8cm}

Spin Muon Collaboration (SMC)
}

%\vskip 1.0cm
 
\author{
B.~Adeva$^{20}$,
E.~Arik$^{2}$,
A.~Arvidson$^{23,u}$,
B.~Bade\l ek$^{23,25}$,
G.~Baum$^{1}$,
P.~Berglund$^{8}$,
L.~Betev$^{13,o}$,
R.~Birsa$^{22}$,
N.~de~Botton$^{19}$,
F.~Bradamante$^{22}$
% ...
}

\vspace{1.5cm}

\affiliation{
\mbox{$\,$University of Bielefeld, Physics Department,
                33501 Bielefeld, Germany }\\
\mbox{$\,^{2}$ Bogazi\c{c}i University and Istanbul Technical University,
		 Istanbul, Turkey }\\
\mbox{$\,^{3}$ University of Bochum, Physics Department, 44780 Bochum, Germany }\\
\mbox{$\,^{4}$ University of California, Department of Physics,
                Los Angeles, 90024~CA, USA }\\
\mbox{$\,^{5}$ CERN, 1211 Geneva 23, Switzerland}\\
\mbox{$\,^{6}$ University of Freiburg, Physics Department,
                79104 Freiburg, Germany }\\
\mbox{$\,^{7}$ GKSS, 21494 Geesthacht, Germany }\\
\mbox{$\,^{8}$ Helsinki University of Technology, Low Temperature
                Laboratory and Institute of Particle Physics Technology,
                Espoo, Finland}\\
\mbox{$\,^{9}$ University of Houston, Department of Physics, and Institute for Beam Particle Dynamics,
                Houston, 77204 TX, USA }\\
\mbox{$\,^{10}$ JINR, Dubna, RU-141980 Dubna, Russia}\\ 
\mbox{$\,^{11}$ University of Mainz, Institute for Nuclear Physics, 55099 Mainz, Germany }\\
\mbox{$\,^{12}$ University of Mons, Faculty of Science, 7000 Mons, Belgium}\\
\mbox{$\,^{13}$ University of Munich, Physics Department, 80799 Munich, Germany }\\
\mbox{$\,^{14}$ Nagoya University, CIRSE and Department of Physics, Furo-Cho,
                 Chikusa-Ku, 464 Nagoya, Japan }\\
\mbox{$\,^{15}$ NIKHEF, Delft University of Technology, FOM and Free University, 
                1009 AJ Amsterdam, The Netherlands }\\
\mbox{$\,^{16}$ Northeastern University, Department of Physics, Boston, 02115 MA, USA }\\
\mbox{$\,^{17}$ Northwestern University, Department of Physics, Evanston, 60208 IL, USA }\\
\mbox{$\,^{18}$ Rice University, Bonner Laboratory, Houston, 77251-1892 TX, USA }\\
\mbox{$\,^{19}$ C.E.A.~Saclay, DAPNIA, 91191 Gif-sur-Yvette, France }\\
\mbox{$\,^{20}$ University of Santiago, Department of Particle Physics, 15706 Santiago de Compostela, Spain }\\
\mbox{$\,^{21}$ Tel Aviv University, School of Physics, 69978 Tel Aviv, Israel }\\
\mbox{$\,^{22}$ INFN Trieste and University of Trieste, Department of Physics, 34127 Trieste, Italy}\\
\mbox{$\,^{23}$ Uppsala University, Department of Radiation Sciences, 75121 Uppsala, Sweden}\\
\mbox{$\,^{24}$ University of Virginia, Department of Physics, Charlottesville, 22901 VA, USA }\\
\mbox{$\,^{25}$ So\l tan Institute for Nuclear Studies and Warsaw University,
                 00681 Warsaw, Poland }\\
\mbox{$\,^{26}$ Yale University, Department of Physics, New Haven, 06511 CT, USA }\\ 
}

\altaffiliation{
$\dagger$ Deceased\\
$\,^{a}$ Now at CERN, 1211 Geneva 23, Switzerland\\
$\,^{b}$ Now at Texas Technical University, Lubbock TX 79409-1051, USA\\
$\,^{c}$ Now at CIEMAT, Avda Complutense 22, 28040, Madrid, Spain\\
$\,^{d}$ Now at Fermi National Accelerator Laboratory,
               Batavia, 60510 Illinois, USA\\
$\,^{e}$ Now at University of Wisconsin, USA \\
$\,^{f}$ Now at NIKHEF, 1009 AJ Amsterdam, The Netherlands\\
$\,^{g}$ Now at Bristol University, Bristol, UK\\
$\,^{h}$ Now at Brookhaven National Laboratory,Upton, 11973 NY, USA\\
$\,^{ih}$ Now at Dept. of Physics and Astronomy, SUNY at Stony Brook, Stony Brook, NY 11974, USA \\
$\,^{jh}$ Now at Univ. of Illinois at Urbana-Champaign, 405 North Mathews Av. Urbana, Illinois 61801, USA \\
$\,^{i}$ Permanent address: Miyazaki University, Faculty of Engineering,
               889-21 Miyazaki-Shi, Japan\\
$\,^{j}$ Permanent address: Paul Scherrer Institut, 5232 Villigen,
                   Switzerland\\
$\,^{k}$ Permanent address: The Institute of Physical and
               Chemical Research (RIKEN), wako 351-01, Japan\\
$\,^{l}$ Permanent address: University of California,
                 Institute of Particle Physics,
                 Santa Cruz, 95064 CA, USA\\
$\,^{m}$ Permanent address: KEK, Tsukuba-Shi, 305 Ibaraki-Ken, Japan\\
$\,^{n}$ Now at SBC Warburg Dillon Read, CH-4002 Basel, Switzerland  \\
$\,^{o}$ Now at University of California, Department of Physics,
                Los Angeles, 90024~CA, USA \\
$\,^{p}$ Permanent address: University of Buenos Aires,
               Physics Department, 1428 Buenos Ai\-res, Argentina \\
$\,^{q}$ Permanent address: Rice University, Bonner Laboratory,
                Houston, TX 77251-1892, USA\\
$\,^{r}$ Permanent address: Warsaw University of Technology, 00-665 Warsaw, Poland\\
$\,^{s}$ Now at F.Hoffmann-La Roche Ltd., CH-4070 Basel, Switzerland\\
$\,^{t1}$ Now at Oak Ridge National Laboratory, Oak Ridge,
               TN 37831-6393, USA\\
$\,^{u}$ Now at The Royal Library, 102 41 Stockholm, Sweden\\
$\,^{v}$ Now at Old Dominion University, Norfolk, VA 23529, USA\\
$\,^{w}$ Now at  University of Bonn, 53115, Bonn, Germany\\
$\,^{x}$ Now at  University of Illinois at Urbana-Champaign, USA\\
$\,^{y}$ Now at  SUBATECH, University of Nantes, UMR IN2P3/CNRS, 44307, Nantes, France\\
$\,^{z}$ Now at  Lawrence Berkeley National Laboratory, Berkeley, CA 94720, USA \\
}

%\collaboration{Spin Muon Collaboration}\noaffiliation
\vspace{3cm}

\begin{abstract}

We present a measurement of the longitudinal spin cross section asymmetry
for deep inelastic muon-nucleon interactions with two high
transverse momentum hadrons in the final state.
Two methods of event classification are used to increase the contribution of
the Photon Gluon Fusion process to above   30\%. The  most effective one,
based on a  neural network approach, provides the asymmetries
$A_p^{\ell N \rightarrow \ell h h X} =
0.030 \pm 0.057 \pm 0.010$ and $A_d^{\ell N \rightarrow \ell h h X} =
0.070 \pm 0.076 \pm 0.010$.
From these values we derive  an averaged gluon polarization
 ${{\Delta G} /  {G}}= -0.20\pm$0.28$\pm$0.10 at an average
fraction of nucleon momentum carried by gluons $\langle \eta \rangle = 0.07$.

\end{abstract}


\maketitle

% body of paper here

\section {Introduction}
 
The Spin Muon Collaboration (SMC) has extensively studied  polarized deep 
inelastic lepton-nucleon scattering using the high energy muon beam at CERN 
and large polarized hydrogen and deuterium targets. This program was 
initiated by the observation in a previous CERN experiment (EMC) that only 
a small fraction of the proton spin is carried by the spin of 
the quarks~\cite{EMC}. The SMC results have confirmed this observation
for protons and provided the first measurement of the spin structure of 
deuterons which allowed for the verification of the fundamental Bjorken sum 
rule~\cite{smc_d92,smc_final}.  

The high energy polarized data from SMC, combined with the high precision data
from the SLAC~\cite{slac} and DESY~\cite{hermes_g1} experiments at lower energy, 
cover a kinematic range
allowing for a QCD analysis of the spin structure function $g_1$.
Various analyses have been performed at next-to-leading order with different
input parameterizations for the polarized parton densities and different 
choices of the fitted parameters~\cite{qcd_exp,qcd_th}. 
They provide consistent results for the polarized quark densities but bring 
little information on the polarized
gluon density $\Delta G$. This is an expected feature since $g_1$ is sensitive
to gluons only through its $Q^2$ evolution and the available $g_1$ data cover 
only a narrow range in $Q^2$ at a given value of $x$. In particular it is still
not possible to test the hypothesis, formulated many years ago,
that the gluon spin may account for a sizable fraction of the nucleon 
spin~\cite{altarelli}. 

A direct measurement of the gluon polarisation is possible via the Photon 
Gluon Fusion (PGF) 
process, which is illustrated in Fig.\ref{fig:pgf}, together with the two 
other lowest order diagrams: the virtual photon absorption
(leading process "LP" ) and gluon radiation (QCD Compton scattering "QCD-C"). 
Since the contribution of the PGF diagram is small, the event selection 
procedure should be very effective in discriminating the
PGF process from other channels. This can
be achieved either by selecting events where a charmed particle is produced 
(e.g. a D meson) or events with hadrons of large transverse momenta ($p_T$) 
relative to the virtual photon 
direction~\cite{carlitz,pt1}.
Both possibilities will be used in the COMPASS experiment presently running at 
CERN~\cite{compass}. 

In this paper we present an evaluation of the gluon polarization, 
$\Delta G/G$, from the SMC data. We limit the analysis to  the DIS region 
($Q^2>1$~GeV$^2$) and  select events with high  $p_T$  hadrons.   
The SMC experimental setup was not optimized for the detection of hadrons 
produced at large angles, so the precision of the result is obviously limited. 
This is, however, the first attempt to tag PGF with light quark production 
in a DIS experiment.  

A determination of the gluon polarization from events with high $p_T$ hadrons 
has been attempted on the ep data from the HERMES experiment~\cite{pt_hermes}
at lower incident energy and in a kinematic range where quasi-real 
photo-production is dominant.

\section{Formalism}

Experimentally observed   spin-dependent effects   are small and have to be 
determined from the cross section asymmetry defined as the ratio of polarized 
($\Delta \sigma$) and unpolarized  ($\sigma$) cross sections
\be
 A^{\ell N}={{\Delta \sigma } \over {2 \sigma}}=
{{\sigma^{\uparrow \downarrow} - \sigma^{\uparrow \uparrow}} \over 
{\sigma ^{\uparrow \downarrow}+ \sigma^{\uparrow \uparrow}}},
\label{eq:asym}
\ee
where $\uparrow \downarrow$  and $\uparrow \uparrow$  refer to  anti-parallel 
and parallel configurations of the nucleon and  incoming lepton spins. 
At the parton level the hard scattering cross section  consists of 
three terms corresponding to the LP, QCD-C and PGF processes. 
According to the factorization theorem  $\sigma$ and $\Delta\sigma$
 can be written as convolutions of the
parton  distributions ($F$, $\Delta F$), hard-scattering cross sections  
($ \hat{\sigma}, \Delta\hat{\sigma}$) and fragmentation functions $D$ 
of partons into hadrons:

\begin{center}
$ \sigma =  F \otimes \hat{\sigma} \otimes D $
\end{center} 
\be
\Delta \sigma = \Delta F \otimes \Delta \hat{\sigma} \otimes D .
\label{eq:unpol}    
\ee
The parton distributions  stand for quarks, antiquarks,  and gluons. 
The spin dependent distributions are denoted 
$\Delta q=q^{\uparrow} - q^{\downarrow}$ for quarks, antiquarks and  
$\Delta G =G^{\uparrow} - G^{\downarrow}$
for gluons and  the  corresponding spin-averaged ones 
$q=q^{\uparrow} + q^{\downarrow}$
and   $G =G^{\uparrow} + G^{\downarrow}$.
Here, the up and down arrows correspond to parallel and anti-parallel   
configurations  of the parton and nucleon spins.

After insertion of the full expression for $\sigma$  and $\Delta \sigma$ 
into Eq.~(\ref{eq:asym}), 
the final expression for the cross section asymmetry  
with production of at least  two hadrons with large transverse momenta, 
$A^{\ell N \rightarrow \ell hhX}$, reads
%\be
\begin{eqnarray}
 A^{\ell N \rightarrow \ell hhX} &=&
\nonumber \\
  {{\Delta q} \over {q}}&&( \langle\hat{a}_{LL} \rangle^{LP}  R_{LP} +
  \langle\hat{a}_{LL}\rangle^{QCD-C}  R_{QCD-C} ) + 
\nonumber \\
  {{\Delta G} \over {G}}&&\langle\hat{a}_{LL}\rangle^{PGF} R_{PGF},
\label{eq:gluon}
%\ee
\end{eqnarray}
where   
$\langle\hat{a}_{LL}\rangle= {\langle{\Delta \hat{\sigma}}/
{2\hat{\sigma}}\rangle}$ 
are the average partonic asymmetries and   $R$ the cross section ratios
of the different processes shown in Fig.~\ref{fig:pgf},  
with respect to the total cross section in the selected sample.
The asymmetry $A^{\ell N \rightarrow \ell hhX}$  thus permits 
an evaluation of the gluon polarization if all other elements in
Eq.~(\ref{eq:gluon}) are known.
The quark asymmetry $\Delta q / q$ is approximated by
the value of $A_1$ obtained in inclusive measurements.                  
The partonic asymmetries $\hat{a}_{LL}$ are calculated for simulated
events and averaged over the selected sample;
in the kinematic region covered by the SMC data,
they are found to be positive for the first two processes and negative for PGF.
The ratios $R$ are  taken from the simulated sample to which the same selection
criteria are applied as to the data.

The statistical precision of the  gluon polarization determined 
from Eq.~(\ref{eq:gluon}) depends on the 
precision of the  measured asymmetry $A^{\ell N \rightarrow \ell hhX}$ and  
on the fraction of PGF events ($R_{PGF}$)
in the final sample. 
Therefore 
the aim  of the present analysis is to select a large enough sample 
with a maximal contribution of PGF events.

The description of hadron production in DIS muon  data   
in terms of the three processes of Fig.~\ref{fig:pgf} has been successfully 
tested in previous experiments~\cite{emc,e665}.
Other processes, such as those involving resolved photons, are expected 
to have  small contributions for $Q^2$ above 1~GeV$^2$ and are not 
considered here.

 \section{The experiment}

The experimental setup at the CERN muon beam consisted
of three major components: a polarized target, a magnetic spectrometer and
a muon beam polarimeter.
A detailed description of the experiment and of the analysis of
the inclusive data can be found in Refs.~\cite{smc_final,smc_prot}.
The muon beam polarization, $P_B$, was determined from the spin
asymmetries measured in polarized muon-electron scattering and
from the energy spectrum of positrons from muon decays and
was found to be $-0.795 \pm 0.019$ for an average beam energy of 
187.4~GeV~\cite{beam_pol}.
The target consisted of two cells filled with butanol, deuterated butanol or
ammonia~\cite{smc_target}.
The two cells were polarized in opposite directions by
dynamic nuclear polarization.
The average target polarizations, $P_T$, were approximately $0.90$ for protons
and $0.50$ for deuterons, with a relative error $\Delta P_T/P_T$ of~3-5\%.
The polarization was reversed five times a day.

The counting rate  asymmetry, $A^{exp}$, is
determined from the  number of events  counted  in
upstream and downstream target cells before and after polarization reversal.
This is done 
by solving the resulting second order equation, as described in \cite{lucas}.

The cross-section asymmetry, $A^{\ell N \rightarrow \ell hhX}$, is related 
to $A^{exp}$  by:
\begin{equation}
%A^{exp}=P_B P_T f A^{\ell N \rightarrow \ell hhX}
A^{\ell N \rightarrow \ell hhX} = \frac{1}{P_B P_T f} A^{exp} \, ,
\end{equation}
where $f$ is the effective dilution factor, which takes into account the 
dilution of spin asymmetries by the presence of unpolarizable
nuclei in the target
and also by radiative effects on the nucleon.
The effect of unpolarizable materials can be expressed  in terms of
the numbers $n_A$ of nuclei with mass
number $A$ and the corresponding total spin-independent
cross sections $\sigma_A^{tot}$. 
The radiative effects on the nucleon \cite{smc_prot,terad} are taken into
account through the ratio
of one photon exchange to total cross-sections
$\rho = {\sigma_{p,d}^{1\gamma}}/{\sigma_{p,d}^{tot}}$.
The evaluation of the effective dilution factor for inclusive events
and for events with observed hadrons is described in Ref.~\cite{smc_final}.
 Polarized radiative corrections are applied to the asymmetries as described 
in Refs.~\cite{smc_prot,polrad}.
In this analysis polarized radiative corrections and dilution  due to 
radiative effects are reduced because processes without  hadrons  are excluded.

 \section{Sample selection}

The total  sample of data collected by the SMC experiment during 
the years 1993-1996 with muon beam of $E$=190~GeV and longitudinally 
polarized target was used for the analysis. It consists of samples 
of similar size taken on polarized protons  and deuterons.

The standard  cuts on inclusive kinematic variables \cite{smc_final}, 
$\nu=E-E'>$15~GeV
and $E^{'}>$19~GeV were imposed to reject events with poor kinematic
resolution and muons from hadron decay, respectively.
The cut  $y=\nu/E<$0.9
removes a   region where the uncertainty due to radiative corrections 
becomes  large. Two other cuts were applied in close relation to 
the formalism used in the analysis:
a cut $Q^{2}>$1~GeV$^2$ rejects the region dominated by nonperturbative 
effects and allows  to interpret the results in terms of partons.
A cut $y>$0.4 removes events which carry little spin information
due to a small virtual photon polarization. In addition, cuts on the muon 
scattering angle were applied in order to match the angular acceptance of 
the hardware triggers.

In the leading  process (LP) most hadrons have small $p_T$ as only the 
intrinsic $k_T$ of quarks in the nucleon~\cite{kt} and the fragmentation 
mechanism contribute to it.
A different  situation occurs for QCD-C and PGF, where
hadrons mainly acquire transverse momentum from primarily  produced partons.
For this reason, the requirement of two observed hadrons with large
transverse momenta enhances the contribution of the PGF
and QCD-C processes in the selected sample. 

In the present  analysis, the  events of interest include a reconstructed beam 
muon, a scattered muon, and at least two charged hadrons.
They  represent   about 20\% of the total number of events with reconstructed
beam and scattered muons, used for inclusive studies.
Hadron tracks were accepted if they could be  associated to the primary 
interaction point, i.e. the vertex, defined by  the beam and scattered muons.
The same association criteria  as
in the SMC analysis of Ref.~\cite{smc_final} were applied.
In order to suppress  the contribution from the target fragmentation region,
cuts on the reduced longitudinal momentum of the hadron, $x_F > 0.1$, and
on the hadron fractional energy, $z=E_h/\nu > 0.1$, were applied.

The further requirement of two hadrons with $p_T >0.7$~GeV selects
about~5\% of the events passing  all previous cuts.
The electron contamination to this sample is expected to be negligible because
electrons are generally produced at low $p_T$. This is confirmed by the ratio
of the energy deposited in the electromagnetic part of the calorimeter to the
total deposited energy, which does not show any peak at 1.0 for tracks 
with $p_T > 0.5$~GeV.
After all selections the total number of  remaining events amounts to  
about 80k for the proton and  70k for the deuteron sample.

\section{Monte Carlo simulation}

\subsection{Conditions for MC generation}

The interactions  were simulated  using the LEPTO 6.5 generator~\cite{lepto} 
with a leading order parameterization of the unpolarized parton 
distributions~\cite{GVR-94}. 
The spin dependent effects were calculated using POLDIS~\cite{poldis} with 
a consistent set of polarized parton distributions~\cite{gs96}.
The kinematic limits of the MC generation were defined so as to cover the full 
kinematic region of the data. Default values were used for  most of 
the steering parameters of the LEPTO  generator.
Below we discuss only the modified  conditions  and parameters.

The matrix  elements of first order QCD processes  exhibit collinear  
divergences in the cross channel and different schemes are used to avoid  
such singularities. 
The so-called $z\hat{s}$ scheme, which allows for lower values of 
the $\gamma ^{*}$-parton center of mass energy $\sqrt{\hat{s}}$,
was used in the simulation with  modified cut-off parameters. 
The effect of the cut-off values on any observable distribution
for events with high $p_T$ hadrons 
is only marginal. 

The description of interactions requires the choice of two scales:
a factorization scale, which appears in the parton densities, 
and a renormalization scale which appears  in expressions depending on  
the  strong coupling constant~$\alpha_{s}$.  
Here the usual  choice of $Q^2$ was  made in both cases.
 In   these conditions, after kinematic cuts on event variables only, the generated sample 
contains 8\%  PGF events.

In order to describe the data, it was found necessary to change the values of
two fragmentation parameters in JETSET~\cite{jtst}.
The function 
$f(z) = z^{-1}(1-z)^{a} e^{-bm^{2}_{T} / {z}}$, where $m_{T}^2=m^2+p_{T}^2$ 
and $m$ is the mass of the quark, expresses the probability that a fraction 
$z$ of the available energy will be carried away by a newly created hadron. 
The parameters ($a$, $b$) were modified from their default values (0.3, 0.58) 
to (0.5, 0.1), a change making the fragmentation softer.
This modification was inspired by a similar study done by the  
HERMES experiment~\cite{naomi,phd_frag}  and  seems to work also in 
the present case, with smaller deviations from the default values.
 However, we are looking  here at a  particular  sample  and  have no   
possibility to check if  the Monte Carlo sample generated with these  
modifications would correctly describe the full data.
The uncertainty connected with these  modifications  has been estimated    
and included in the systematic error.

\subsection{Simulation of experimental  conditions} 

 The scattered muon track of each simulated event was followed
through the magnet aperture. Trigger conditions were checked and prescaling
factors applied in order to reproduce the relative trigger rates in 
the simulated sample. Kinematic smearing was applied to muon and hadron tracks 
and geometric smearing  to the vertex position. In addition, the loss
of tracks due to chamber inefficiencies was taken into account by
applying detector plane efficiencies to the simulated events and by removing 
the tracks which did not fulfill the minimal requirements for reconstruction.    

Secondary interactions of hadrons have to be taken into account
to reproduce  the distribution of interaction vertices along the target axis. 
Hadrons were rejected from the sample according to the probability of  
re-interaction  in the polarized target material.
As an example, Fig.~\ref{fig:vertex} shows
the agreement obtained for the vertex position along the beam axis 
in one of the proton data sets. 

The simulation was performed for each year of data taking separately.
To  get a  good description of the  kinematic variables it was required  to use 
specific   beam parameters for every year, including small changes in angles, 
and  to  take into account the exact target position.

\subsection{Comparison  of simulations and  data} 

The distributions of kinematic variables as well as the particle distributions  
in detectors were checked with identical selection criteria applied to data and
 MC. For the simulated events  the cuts were applied to the smeared 
variables. The distributions for data and MC were normalized to the same 
number of events. The distributions of $x$ and $Q^2$ for  interactions on 
protons are presented in Fig.~\ref{fig:kinem}.  
The obtained  agreement is  at the level of 10-25\% for all kinematic event  
variables. The level of agreement for deuterons is  very similar~\cite{kk_phd}.
 
The same comparisons were done for hadron variables. For simulations performed 
with the unmodified fragmentation function clear discrepancies are observed 
for the hadron production angle  $\theta$ and the  longitudinal momentum $p_L$,
while satisfactory agreement is obtained for  $p_T$, except at the highest 
values.  The observed differences at the highest 
values of $p_T$ can be explained by the approximate  description of 
the non-Gaussian tails of the distributions used for smearing and by 
the effects of real photon radiation, which are not taken into account 
in the present analysis.
It was checked that the discrepancy for  the   $\theta$ angle could not  
be removed by using different smearing parameterizations or even by 
an artificial increase of smearing.
Agreement between data and simulation could only be achieved by applying 
a cut on the hadron production 
angle $\theta > 0.02$~rad. This cut, however,  removes about 25\% of 
the selected sample and cannot be justified since there is no reason  
why the simulation should not describe the hadrons produced at low $\theta$.  
Therefore modified simulation conditions providing a  better description 
of the data were searched for.

When the modifications of  the fragmentation function parameters are applied 
(cf. Section V.A), the agreement  becomes satisfactory over a wide range 
of $\theta$ and  $p_L$. The comparison of the $p_L$ and $\theta$ distributions  
is shown in Fig.~\ref{fig:modhadrons} for  the hadron with highest $p_T$. 
The second hadron is also well described by the MC~\cite{kk_phd}.
We concluded that the parameters of the  longitudinal fragmentation function  
$f(z)$ have to be modified in order to obtain a good description of the data 
over the full range of hadron production angle $\theta$.
Since it is difficult to check if the modified set of parameters  correctly 
describes the semi-inclusive hadron distributions,
the  analysis has  been  performed in parallel with modified fragmentation
as well as with the standard fragmentation and an additional cut on 
$\theta > 0.02$~rad.

\section{Selection of the PGF process}

In order to compare the merits of various selections of PGF events, we will
use the {\it efficiency} $\epsilon$, which is the ratio of the number of PGF
events accepted by the selection criteria to the total number of PGF events,
and the {\it purity} $R_{PGF}$ (Eq.3), which is the ratio of the number 
of selected PGF events to the total number of selected events. The optimal 
selection is obviously the one providing the highest values of $\epsilon$ 
and $R_{PGF}$ but, in general, an increase of the former will result in 
a decrease of the latter.

The purity is  $ 0.11$ for the full sample of events with at least 2 charged 
hadrons. The additional requirement of two  hadrons with $p_T>0.7$ GeV defines 
our reference sample for which
$R_{PGF} = 0.24$ and, by definition,  $\epsilon = 1$. 

The effects of cuts were studied  for  the following  variables: 
$p_{T1}$, the sum $p_{T1}^{2}+p_{T2}^{2}$, hadron charges 
(same or opposite sign), the azimuthal angle $\phi$ between the momenta
of the two hadrons with respect to the virtual photon direction, 
and the invariant mass of the two hadrons (see also Ref.~\cite{hubert}). 
It was found that the selection on $\sum p_T^2$ is optimal for enhancing 
the PGF purity and that further requirements on the hadron charges do not
bring any significant improvement. Fig.~\ref{fig:comp_traf} shows the variation
of $R_{PGF}$ with $\epsilon$  when the cut on $\sum p_T^2$ is varied up to 
4~GeV$^2$.
It is seen that the purity increases only very slowly when the cut is made 
more restrictive while the efficiency drops very rapidly. This can be 
understood by the fact that one of the background processes (QCD-C) has 
a similar dependence on  the $\sum p_T^2$ cut as PGF.
The approximation made in Eq.~(\ref{eq:gluon}) by the use of $A_1$ for 
the asymmetry on quarks is only valid if the fraction of PGF events in 
the selected sample is much higher than in the inclusive one, i.e. close to 
the maximum value of 0.33. The efficiency also needs to be sufficiently high 
to allow a meaningful analysis. As a compromise, we have fixed the cut 
at 2.5 GeV$^2$, which corresponds to $\epsilon$ = 0.30 and $R_{PGF}$ = 0.31.

The combination of  several  variables into a single parameter has also been
investigated in a classification procedure based 
on a neural network \cite{acta_paper,kk_phd}.
We considered the variables which characterize the DIS event ($x$, $Q^2$, 
$y$, and the multiplicity of tracks)
and those which describe the two selected hadrons with highest $p_T$ 
(transverse and longitudinal hadron momenta, charges of the hadrons, 
energy fraction of the hadrons, and the azimuthal angle  
$\phi$).
The classification procedure was trained on a Monte Carlo sample where the
actual process is known for each event.
As a result, the procedure provides a single value, called "NN response", 
within the range (0,1). 
High values of this  response correspond to events which, according to 
the classification algorithm, are more likely to be PGF than background 
processes.  A threshold on the network response can thus be used 
to select a PGF enriched sample.

The variation of $R_{PGF}$ vs. $\epsilon$  for various choices of 
the NN response threshold is also shown in Fig.~\ref{fig:comp_traf}. 
It is observed that at equal efficiency the NN approach always provides
samples with higher purity than the selection based on $\sum p_T^2$. 
For further analysis, a threshold of 0.26 was chosen, which corresponds 
to $ R_{PGF} = 0.33$ and $\epsilon = 0.56$.
A similar  purity is obtained with the $\sum p_T^2$ cut at 2.5 GeV$^2$ but 
with an efficiency of 30\%.                        
Therefore a better statistical precision on the measured asymmetry will be 
obtained with the neural network method.
Alternatively, a higher NN threshold corresponding to a PGF efficiency of 30\%
would yield a sample where the purity is about 37\%, i.e. 6\% higher than 
the value obtained with the $\sum p_T^2$ cut.
The comparison of the two selected samples shows that the NN procedure 
selects a large fraction of events with $\sum p_T^2 > 2.5$ GeV$^2$ but also 
covers the lower range of $\sum p_T^2$. It was also checked that 
the distributions of NN responses are compatible for data and 
Monte-Carlo events. 
  
\section{Spin asymmetries  $A^{\ell N \rightarrow \ell hhX}$}
  
The SMC data taken from 1993 to 1996 were split into periods of data taking, 
corresponding to about 15 days each. The asymmetry for a given  year is 
the weighted average  of the asymmetries calculated for each  period of data 
taking. Splitting the data into smaller subsamples gives identical results
within the expected statistical fluctuations.  
The distribution of the vertex position along the beam axis, as presented 
in Fig.~\ref{fig:vertex}, shows  that the ratio of acceptances 
for the  upstream to downstream target cells is about  0.7.
The method used for asymmetry calculation, described in \cite{lucas}, 
is suited for such an acceptance difference. 

The asymmetry calculations were done for the entire sample which has a purity 
$R_{PGF} = 0.24$ and for the two selection methods with enhanced $R_{PGF}$ 
($\sum p_T^2 > 2.5$ GeV$^2$ and NN response $>$ 0.26).
The results given in Fig.~\ref{fig:asym_vs_pt2} and Table~\ref{tabasy} show 
that the asymmetries do not change significantly with the selection. 
Also the asymmetries obtained for proton and deuteron are compatible within 
errors. The statistical error is larger for the selection based on $\sum p_T^2$
because a  smaller fraction of events is selected (28 \% vs. 42 \%).

The errors of the measured  $A^{\ell N \rightarrow \ell hhX}$ asymmetry for 
the selected samples are dominated  by statistics.
The contributions to the systematic uncertainty on 
$A^{\ell N \rightarrow \ell hhX}$  are detailed in Table \ref{data:selfinall} 
for the two selections with enhanced $R_{PGF}$.
The  most  significant  ones come from the false asymmetries, the fraction 
of radiative processes ($\rho$) and the polarized radiative corrections.  
For the false asymmetries an upper limit from the time variation of 
the acceptance was taken under the assumption that the reconstruction for 
each of the three tracks (scattered muon and two hadrons) is affected 
independently.
The method used for estimating these effects is described in Ref.\cite{hubert}. 
The radiative corrections are small due to the limited phase space for 
real photon emission in events where a significant 
fraction of the available energy is taken by the two hadrons with large $p_T$.
The  uncertainties in $\rho$ and polarized radiative corrections were taken 
equal to the full size of the inelastic contribution. 
The effect of real photon radiation on the event kinematics and, in particular,
on the value of $p_T$ itself has not been taken into account in view of 
the limited precision of the present analysis.

\section{Determination of the  gluon polarization} 

 The gluon polarization is evaluated from Eq.~(\ref{eq:gluon}) using 
the measured  $A^{\ell N \rightarrow \ell hhX}$ asymmetry, obtained for 
the samples with enhanced $R_{PGF}$, quoted in Table 1. 
In view of the strong dependence of the resulting gluon polarization on the 
information obtained from the Monte Carlo, special attention was given to 
the agreement of data and simulated events (Figs.2-4).  

  The asymmetry  $A_1(x)$  for each  event is taken from a fit to all 
experimental data and averaged for the full proton and deuteron samples.
The  partonic asymmetries  $\hat{a}_{LL}$ for each  sub-process are calculated 
for each Monte-Carlo event  and averaged.
Their  averaged values for LP and QCD-C are very similar for the two 
selections  namely,
 $\langle\hat{a}_{LL}\rangle^{LP} = 0.8$ and  
 $\langle\hat{a}_{LL}\rangle^{QCD-C} = 0.6$.      
The values for PGF are
  $\langle\hat{a}_{LL}\rangle^{PGF}= -0.44$  and $-0.49$ 
 for the  $\Sigma p_T^2$ cut  and the NN selection, respectively.   
After  selection on $\Sigma p_T^2$ the final proton sample  consists of 
26\% LP, 43\% QCD-C and 31\% PGF, while for  the neural network the fractions 
are  $R_{LP} = 38$\%, $R_{QCD-C} =29$\%  and $R_{PGF} =33$\%.
The contributions  of different processes for the proton and deuteron samples 
differ by less than 2\%.  

The  gluon polarization      
is determined for the   kinematic region covered  by the selected sample and 
corresponds to a given fraction of nucleon momentum carried by gluons $\eta$:
\begin{equation}
\eta = x ( {\hat{s} \over Q^{2}}+1). 
\end{equation}
This quantity is known for simulated events  but cannot be directly determined 
from the data. 
Nevertheless, $\hat{s}$  can be approximately calculated from the virtual 
photon energy in the laboratory system and from the angles 
($ \theta_{1}$, $ \theta_{2}$) defined by the directions of the two hadrons 
with respect to the virtual photon:       
\begin{equation}
\hat{s} \approx \nu^2 {\rm tg} \theta_{1} {\rm tg} \theta_{2}.                  
\end{equation}

To check the validity of  this approximation in our kinematic conditions, 
we have compared the generated $\eta$ 
and the one calculated  from the above equation  for selected PGF events. 
The calculated values are on  average 25\% higher than the generated ones.
The averaged value of the  generated  $\eta$ for the selected PGF events 
in the Monte Carlo is used as the  reference value for  
the result on  $\Delta G/G$.
We have also  checked  the average values  of  $\eta$ calculated  for 
all simulated events  and  obtained  the values  0.15 for the  
cut~$\sum p_T^2 >$~2.5~GeV$^2$ and  0.10 for the NN response~$>$~0.26. 
For both selection methods the values of $\eta$ calculated  for all simulated 
events and for data are very close. The results on the  gluon polarization 
and the  values of $\langle\eta\rangle$ are presented in  Table~\ref{tabgluon}.

In addition to the  systematic errors on the  measured asymmetry  discussed 
in Section~7 and given in Table~\ref{data:selfinall}, 
the asymmetry $A_1$, the fractions $R$, and the partonic asymmetries 
$\langle\hat{a}_{LL}\rangle$  contribute to the systematic error on  
$\Delta G / G$.  
The contribution due to the asymmetry $A_1$ is determined 
from the uncertainty on $A_{1}$  at the averaged value of $x$ and thus
from the errors on the fit parameters. 
The value of $A_1$ at the average $x$  agrees with the average  $A_1$ 
calculated from the fit for each event to within 0.001.

The dominant   contributions to the  systematic error are due to the 
uncertainties on  the values of  $R$ and $\langle\hat{a}_{LL}\rangle$. 
They are  estimated by comparing  the results obtained from Monte Carlo 
simulations with different parameters.    
For this purpose, a sample of LEPTO events  was generated with the same 
kinematic and hadron selections  but with  modified renormalization and 
factorization scales, cut-offs and fragmentation function parameters.
Scales of $Q^2/2$ and $2~Q^2$ were used for comparison and provide an 
estimate of the stability of the leading order approximation used here.
Results with standard and modified parameters (see Section~5.1)
in the fragmentation function were compared.
Since only the  simulations  which reproduce  the data should be considered,
a cut on the hadron angle $\theta$ was applied, as explained in Section~5.3.
The value of the  gluon polarization calculated with this new Monte Carlo  
sample was compared to the one obtained under the conditions described in 
Section 5.1. This procedure was repeated several times with slightly different 
cuts and with different neural network thresholds. 
For the neural network the procedure  is complicated  by the fact that any 
change in the simulation procedure leads to a different selection on the data. 
To avoid the fluctuation of the gluon polarization due to variation of 
the measured asymmetry, the value of this asymmetry was 
artificially frozen when comparing results for different MC samples. 
The individual contributions to the systematic error are given, for both 
selection methods, in Table~\ref{syst:gluon}. It was checked that the effect 
of combined modifications in the Monte Carlo is smaller than the sum of 
the individual uncertainties. The maximal variation of $R_{PGF}$ and 
$\langle\hat{a}_{LL}\rangle$ was found to be 20\% and 4\% respectively.

 As discussed  before, the neural network selection provides a more accurate 
result than the selection based on $\Sigma p_T^2$ cuts. 
However, the statistical error is too large to draw definitive conclusions 
on the contribution of $\Delta G$ to the nucleon spin.  
The systematic uncertainty is small compared  to the statistical error.
The  demand of a good agreement of the simulation with the data sets 
an important limitation on   the estimated  systematic uncertainties.  
For this reason, an increase in statistical precision is expected also to 
lead to further improved systematic uncertainty estimates.

Averaging the results for  proton and deuteron  obtained with the neural 
network classification we obtain  
${{\Delta G} /  {G}} = -0.20 \pm 0.28 \pm 0.10$.

\section {Conclusions}

We have evaluated for the first time   the gluon polarization from
the spin asymmetries measured in lepton-nucleon DIS events with 
$Q^2 > 1$~GeV$^2$ including two hadrons with large  transverse momentum 
in the final state. 
The analysis is performed at leading order in QCD and based on the comparison
of selected data samples with simulated events provided by the LEPTO generator.
The partonic asymmetry $\hat{a}_{LL}$ is mostly negative for
the photon-gluon fusion process while it is positive for
the two competing processes, leading process and gluon radiation.
The relative contribution of photon-gluon fusion is  enhanced
to about 30\% by applying a cut on $\Sigma p_T^2 > 2.5$~GeV$^2$
or by using a neural network classification.
                        
The average gluon polarization obtained for the SMC data is close to zero
with a large statistical error ($\sim 0.30$). The accuracy is limited
by the reduction to  less than 1\% of the DIS sample by the hadron selection 
requirements. It is thus expected to be improved by higher counting rates 
and larger hadron acceptance in ongoing and future experiments.

\section*{ACKNOWLEDGMENT}

This work was supported by Bundesministerium f\"{u}r Bildung, Wissenschaft,
Forschung und Technologie, partially supported by TUBITAK and the Center
for Turkish-Balkan Physics Research and Application (Bogzi\c{c}i University),
supported by the U.S. Department of Energy, the U.S. National Science
Foundation, Monbusho Grant-in-Aid for Science Research (International Scientific
Research Program and Specially Promoted Research), the National Science
Foundation (NWO) of the Netherlands, the Commisariat \`{a} l'Energie Atomique,
Comision Interministerial de Ciencia y Tecnologia and Xunta de Galicia,
the Israel Science Foundation, and Polish State Committee for Scientific
Research (KBN) SPUB no. 134/E-365/SPUB-M/CERN/P-03/DZ299/2000 and
621/E-78/SPB/CERN/P-03/DWM 576/2003-2006 and Grant No. 2/P03B/10725.

\newpage
\begin{widetext}

\begin{figure*}
\begin{center}
\includegraphics[width=13cm]{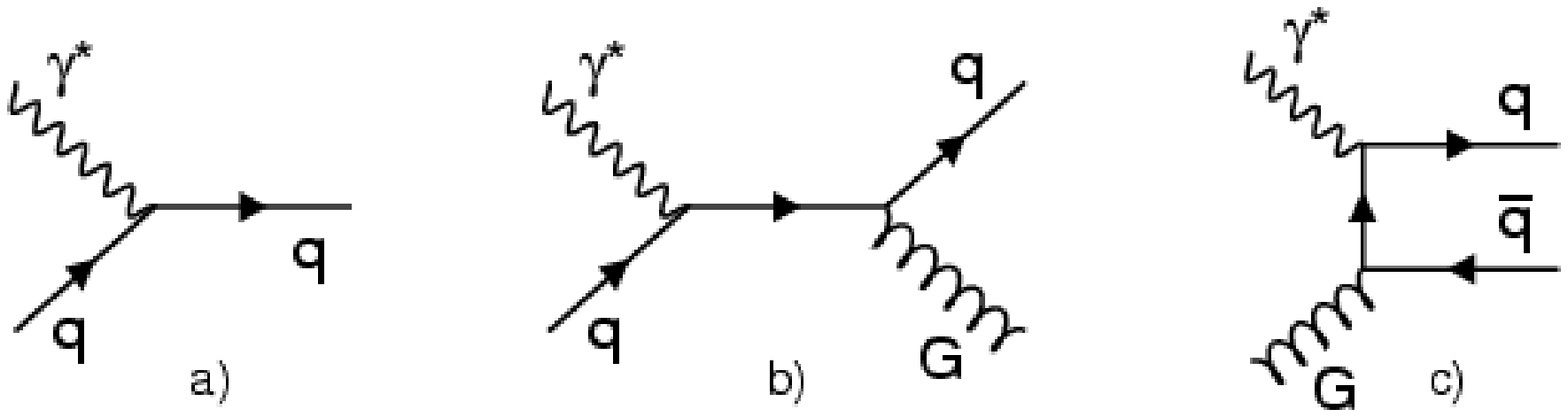}%
\caption{Lowest order diagrams for DIS $\gamma^{\ast}$ absorption: 
a) leading process (LP),
b) gluon radiation (QCD-C), 
c) photon-gluon fusion (PGF).}
\label{fig:pgf}
\end{center}
\end{figure*}

\begin{figure*}
\begin{center}
\includegraphics[width=15cm]{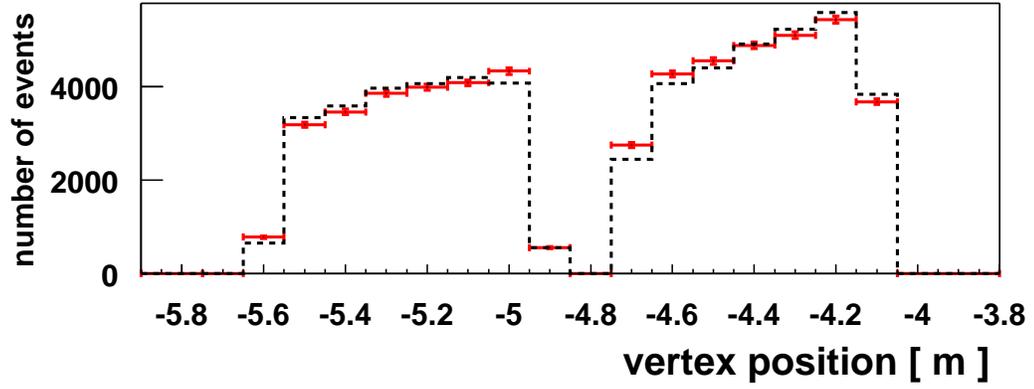}%
\caption{Distribution of vertices along the beam axis.
Points correspond to the proton data from 1993 and the histogram to 
the corresponding MC simulation.}
\label{fig:vertex}
\end{center}
\end{figure*}

\begin{figure*}[b]
\begin{center}
\includegraphics[width=13cm]{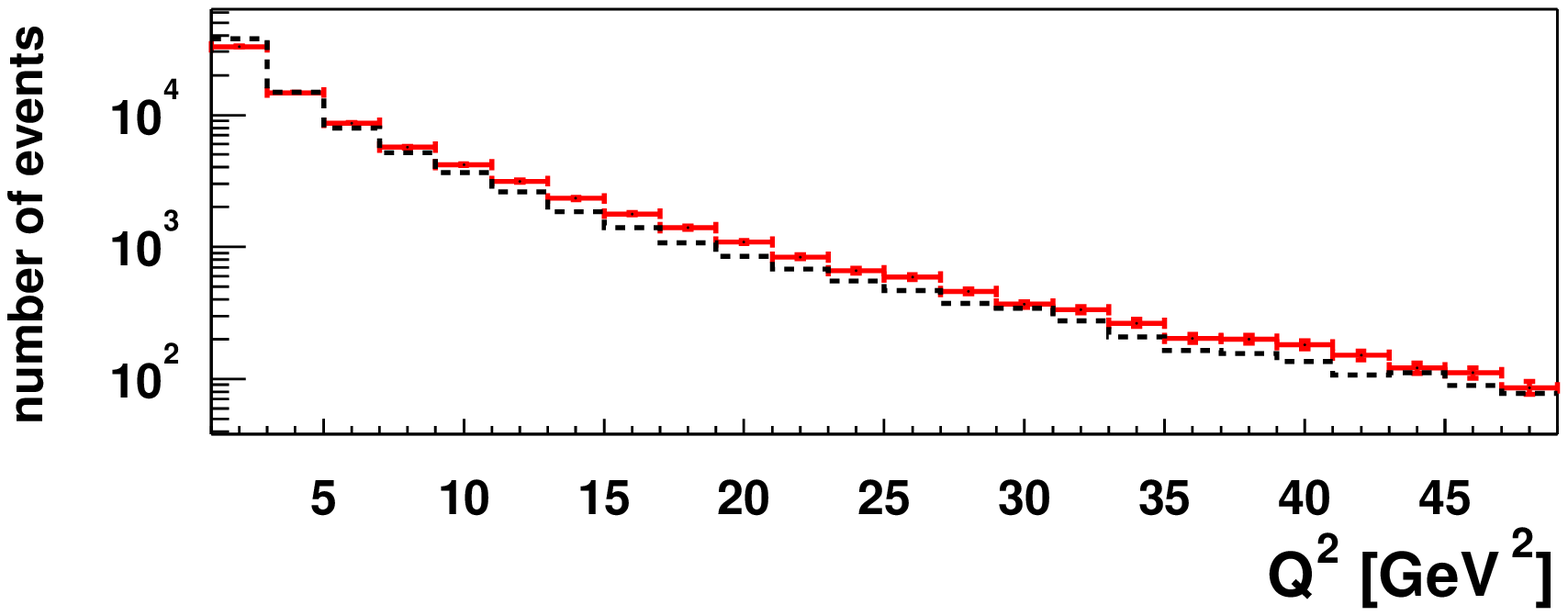}%

\hspace*{.2cm}

\includegraphics[width=13cm]{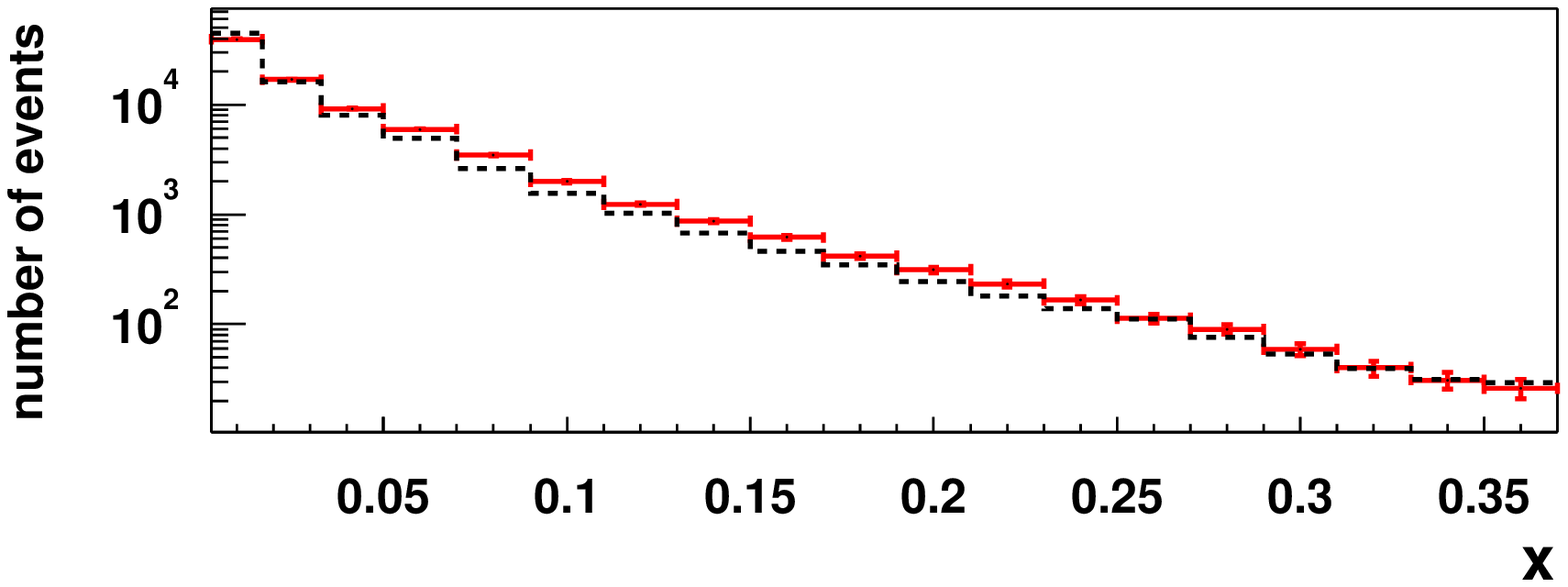}%
\caption{The $x$ and $Q^2$ distributions for the proton case:                  
points correspond to the data and histograms to the Monte Carlo
simulation.}
\label{fig:kinem}
\end{center}
\end{figure*}

\begin{figure*}
\begin{center}
\includegraphics[width=13cm]{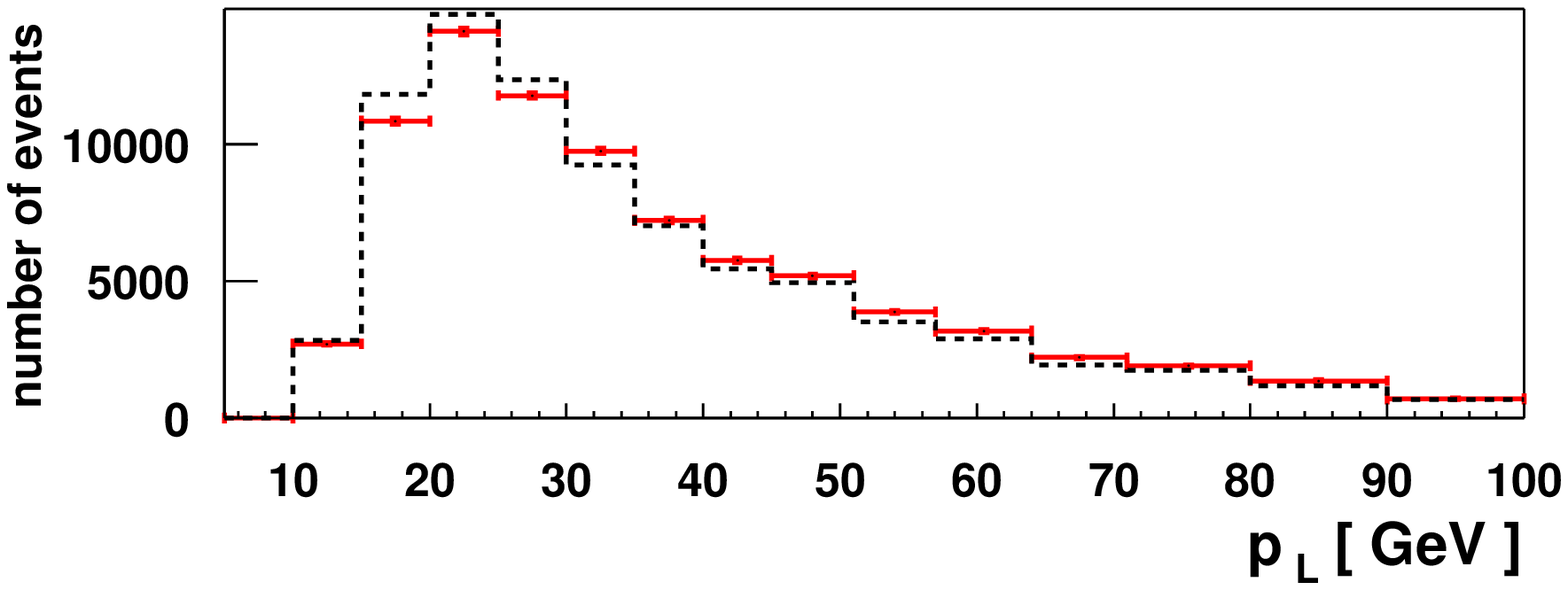}% 

\hspace*{.2cm}

\includegraphics[width=13cm]{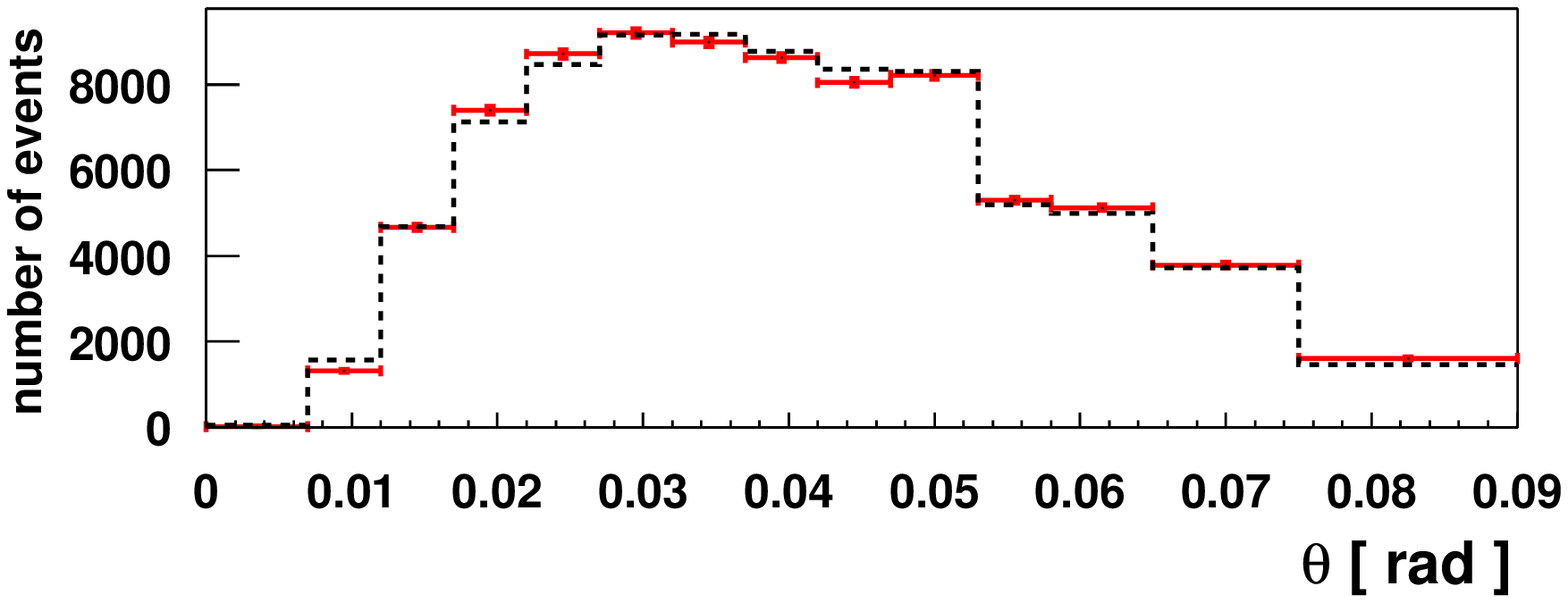}% 
\caption{ Distributions of longitudinal momentum and scattering angle
 for the hadron with the highest $p_T$.
Points correspond to the proton data collected in 1993, 
histograms to the Monte Carlo simulations with the modified 
fragmentation function.}
\label{fig:modhadrons}
\end{center}
\end{figure*}

\begin{figure*}
\begin{center}
\includegraphics[width=14cm]{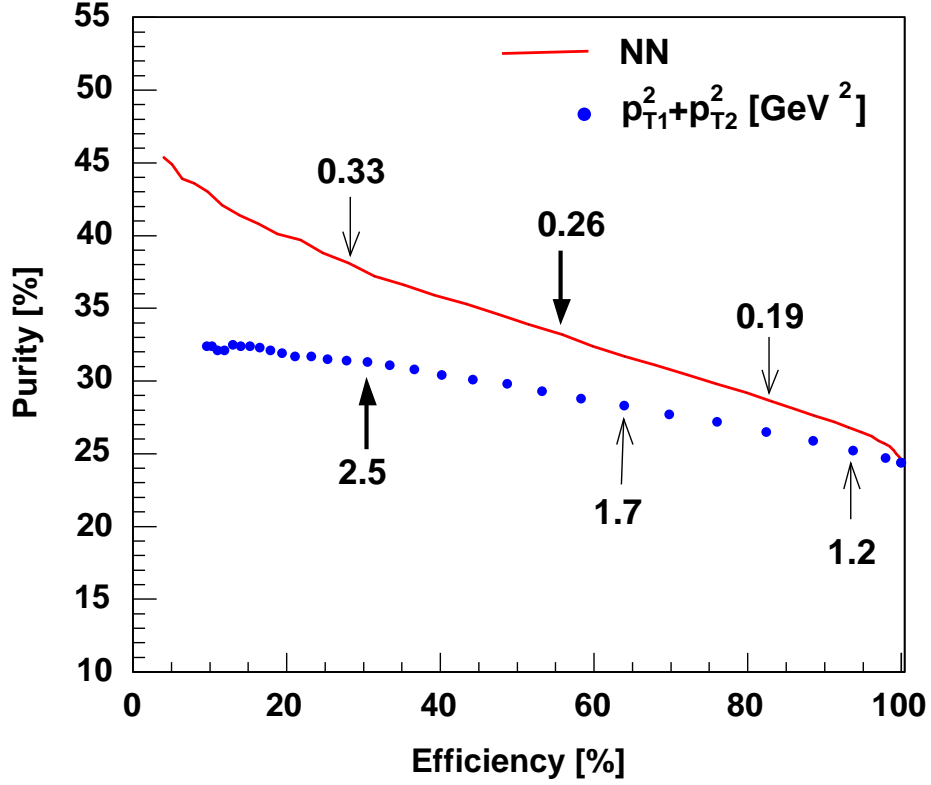}%
\end{center}
\caption{Comparison of purity and efficiency for the selection methods 
based on the cut on $\sum p_T^2$ and the NN response. Simulations correspond 
to the proton sample.}
\label{fig:comp_traf}
\end{figure*}

\begin{figure*}
\begin{center}
\includegraphics[width=14cm]{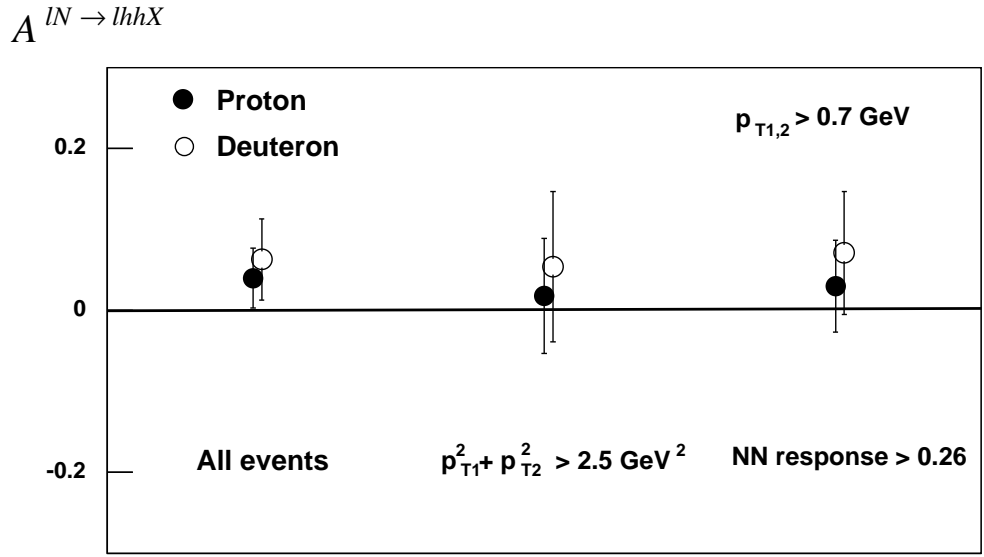}% 
\end{center}
\caption{Measured asymmetry $A^{\ell N \rightarrow \ell hhX}$,  
for proton and deuteron, for events with $p_{T1,2}>0.7~$GeV cut and  
after additional selections on  $\sum p_T^2$ and neural network threshold 
to increase the purity.  }
\label{fig:asym_vs_pt2}
\end{figure*}

\newpage

\begin{table}[t]
\caption{ Measured cross-section asymmetries $A^{\ell N \rightarrow \ell hhX}$ 
for proton and deuteron events with $p_{T1,2}>0.7~$GeV and in the samples 
selected with the  $\sum p_T^2$ cut and with the neural network response 
threshold, each given with statistical  and systematic errors. }
\begin{center}
\begin{ruledtabular}
\begin{tabular}{lcc} \\
\multicolumn{1}{l}{Selection} &  
\multicolumn{1}{c}{$A_{p}^{\ell N \rightarrow \ell hhX}$} &  
\multicolumn{1}{c}{$A_{d}^{\ell N \rightarrow \ell hhX}$}    \\ 
\\\hline \\
 All             &  $0.041$$\pm$$0.037$$\pm$$0.011$  &     
  $0.063$$\pm$$0.050$$\pm$$0.011$  \\
 $\sum p_T^2 > $2.5~GeV$^2$      &  $0.018$$\pm$$0.071$$\pm$$0.010$  & 
   $0.054$$\pm$$0.093$$\pm$$0.008$ \\
 NN response$ > $0.26           &  $0.030$$\pm$$0.057$$\pm$$0.010$  &  
  $0.070$$\pm$$0.076$$\pm$$0.010$ \\ 
\end{tabular}
\end{ruledtabular}
\label{tabasy}
\end{center}
\end{table}

\begin{table}[htb]
\caption{The contributions to the systematic error of 
$A^{\ell N\rightarrow \ell hhX}$
with the $\sum p_T^2 > $2.5~GeV$^2$ cut
and with the neural network response $ > $0.26
for SMC proton and deuteron data.
The first and last contributions are additive; the others  are proportional to
the asymmetry.  }
\begin{center}
\begin{ruledtabular}
\begin{tabular}{l r r r r }  \\
\multicolumn{1}{l}{Contributions to the}
&  \multicolumn{2}{c}{proton data} & \multicolumn{2}{c} {deuteron data}\\

\multicolumn{1}{l}{systematic error on $A^{\ell N\rightarrow \ell hhX}$}
&  \multicolumn{1}{c}{$\Sigma p_T^2$}
& \multicolumn{1}{c} {NN }
&  \multicolumn{1}{c}{$\Sigma p_T^2$}
& \multicolumn{1}{c} {NN } \\
\\\cline{1-3} \hline  \\

 False asymmetries                   & 0.0049  & 0.0049 & 0.0044 & 0.0044   \\
 Target polarization                 & 0.0005  & 0.0008 & 0.0016 & 0.0023   \\
 Beam polarization                   & 0.0007  & 0.0011 & 0.0021 & 0.0029   \\
 Dilution factor \\

    \hspace{5mm} Target composition  & 0.0003  & 0.0001 & 0.0002 & 0.0001   \\

    \hspace{5mm} $\rho$ factor       & 0.0018  & 0.0030 & 0.0054 & 0.0076   \\

 Polarized rad. corr.                & 0.0083  & 0.0083 & 0.0020 & 0.0020   \\
\\
 Total systematic error              & 0.0098  & 0.0102 & 0.0077 & 0.0097   \\

\end{tabular}
\end{ruledtabular}
\label{data:selfinall}
\end{center}
\end{table}

\begin{table}[htb]
\caption{ Gluon polarization for proton and deuteron
for the $\Sigma p_T^2$ cut
and the neural network selection. }
\begin{center}
\begin{ruledtabular}
\begin{tabular}{l c c l  } \\
\multicolumn{1}{c}{Selection } & 
 \multicolumn{1}{c}{$\left({{\Delta G} \over {G}}\right)_{p}$} 
 &  \multicolumn{1}{c}{$\left({{\Delta G} \over {G}}\right)_{d}  $}
 &  \multicolumn{1}{c}{$\langle\eta\rangle$} \\ \\ \hline  \\
 $\sum p_T^2 >$ 2.5~GeV$^2$   
 &  0.11$\pm$0.51$\pm$0.12  &  --0.37$\pm$0.66$\pm$0.12 &  0.09 \\

 NN response$ > $ 0.26        
  & --0.06$\pm$0.35$\pm$0.10  &  --0.47$\pm$0.49$\pm$0.10 & 0.07  \\ \\
\end{tabular}
\end{ruledtabular}
\label{tabgluon}
\end{center}
\end{table}

\begin{table}[htb]
\caption{ Contributions to the systematic error on gluon polarization
for two methods of event selection. }
\begin{center}
\begin{ruledtabular}
\begin{tabular}{l c c } \\
\multicolumn{1}{c}{Source of the uncertainty } & 
\multicolumn{1}{c}{$\Sigma p_T^2$}
& \multicolumn{1}{c}{NN } \\
\\\hline \\

  systematic error \\
 on $A^{\ell N \rightarrow \ell hhX}$   &  0.072(p) 0.057(d)  
  &  0.061(p) 0.063(d) \\

 precision of $A_1$ fit                     & 0.042(p) 0.042(d)  
  &  0.026(p) 0.028(d) \\

 scale change \\
 from $Q^2/2$ to $2~Q^2$   & 0.008  & 0.010   \\

 fragmentation paramr.     & 0.036  & 0.034   \\

 cut-offs in matrix elem.  & 0.015  & 0.008   \\ 

\end{tabular}
\end{ruledtabular}
\label{syst:gluon}
\end{center}
\end{table}

\end{widetext}


\begin{thebibliography}{99}

\bibitem{EMC} EMC, J.~Ashman {\it et al.,} Nucl. Phys. B {\bf 328}, 1 (1989); 
Phys.Lett.B {\bf 206}, 364 (1988).
\bibitem{smc_d92} SMC, B.~Adeva {\it et. al.,} Phys. Lett. B {\bf 302}, 
533 (1993).
\bibitem{smc_final} SMC, B.~Adeva {\it et al.,} Phys.\ Rev. {\bf D58}, 
112001 (1998). 
\bibitem{slac} E142, P.L.~Anthony {\it et al.,} Phys. Rev. D {\bf 54}, 
6620 (1996); \\
E143, K.~Abe {\it et al.,} Phys.  Rev. D  {\bf 58}, 112003 (1998); \\  
E154, K.~Abe {\it et al.,} Phys. Rev. Lett. {\bf 79}, 26 (1997); \\
E155, P.~L.~Anthony {\it et al.} Phys. Lett. B {\bf 458}, 529 (1999).
\bibitem{hermes_g1} HERMES,  
A.~Airapetian {\it et. al.,} Phys. Lett. B {\bf 442}, 484 (1998).
\bibitem{qcd_exp} SMC, B.~Adeva {\it et al.,} Phys.  Rev. D {\bf 58}, 
112002  (1998); \\ 
E155, P.~L.~Anthony {\it et al.,} Phys. Lett. B {\bf 493}, 19 (2000).
\bibitem{qcd_th} G.~Altarelli, R.~D.~Ball, S.~Forte and G.~Ridolfi, 
Nucl. Phys. B {\bf 496}, 337 (1997); \\
E.~Leader, A.~V.~Sidorov and D.~B.~Stamenov, Eur. Phys. J. C {\bf 23}, 
479 (2002); \\
Y.~Goto {\it et al.,} Phys. Rev. D {\bf 62} (2000) 034017; \\
M.~Gl\"uck, E.~Reya, M.~Stratmann and W.~Vogelsang, Phys. Rev. D {\bf 63}, 
094005 (2001); \\
J.~Bl\"umlein and H.~Bottcher, Nucl. Phys. B {\bf 636}, 225 (2002); \\
C.~Bourrely, J.~Soffer and F.~Buccella, Eur. Phys. J. C {\bf 23}, 487 (2002).
\bibitem{altarelli} A.V.~Efremov and O.V.~Teryaev, JINR Report E2-88-287, 
Dubna (1988);\\ 
G.~Altarelli and G.G.~Ross, Phys. Lett. B {\bf 212}, 391 (1988).
\bibitem{carlitz} R.D.~Carlitz, J.C.~Collins and A.H.~Mueller, 
Phys. Lett. B {\bf 214}, 229 (1988).
\bibitem{pt1} A.~Bravar, D.~von Harrach and A.~Kotzinian, 
Phys. Lett. B {\bf 421}, 349 (1998).
\bibitem{compass} COMPASS, G.~Baum {\it et al.},
'COMPASS: A Proposal for a Common Muon and Proton Apparatus for Structure 
and Spectroscopy', CERN-SPSLC-96-14.
\bibitem{pt_hermes}
HERMES, A.~Airapetian {\it et al.,} Phys. Rev. Lett. {\bf 84}, 2584 (2000).
\bibitem{emc} EMC, M.~Arneodo {\it et al.}, Phys. Lett. B {\bf 149}, 
415 (1984); Z. Phys. C {\bf 36}, 527 (1987).
\bibitem{e665} E665, M.R.~Adams {\it et al.}, Phys. Rev. D {\bf 48}, 
5051 (1993); Phys.Rev.Lett.{\bf 72}, 466 (1994). 
\bibitem{smc_prot} SMC, D.~Adams {\it et. al.,} Phys. Rev.  D {\bf 56}, 
5330 (1997).
\bibitem{beam_pol} SMC, B.~Adeva {\it et al.}, Nucl. Instr. 
and Meth. A {\bf 443}, 1 (2000).
\bibitem{smc_target} SMC, B.~Adeva {\it et al.}, Nucl. Instr. and 
Meth. A {\bf 437}, 23 (1999).
\bibitem{lucas} L.~Klosterman, Ph.D. thesis, Delft University of Technology, 
Delft (1995). 
\bibitem{terad}  A.A.~Akhundov {\it et al.}, Fortsch. Phys. {\bf 44}, 373 
(1996); 
Sov. J. Nucl. Phys. {\bf 26}, 660 (1977); {\it ibid.} {\bf 44}. 988 (1986);\\ 
D.~Bardin and N.~Shumeiko, Sov. J .Nucl. Phys. {\bf 29}, 499 (1979).
\bibitem{polrad}  T.V.~Kukhto and N.M.~Shumeiko, Nucl. Phys. B {\bf 219}, 
412 (1983).  
\bibitem{kt} A.~K\"onig, Z. Phys. C {\bf 18}, 63 (1983).
\bibitem{lepto} G.~Ingelman, A.~Edin and J.~Rathsman, Comput. Phys. Commun. 
{\bf 101}, 108 (1997).
\bibitem{GVR-94} M.~Gl\"uck, E.~Reya and A.~Vogt, Z. Phys. C {\bf 67}, 
433 (1995).
\bibitem{poldis} A.~Bravar, K.~Kurek and R.~Windmolders, Comput. Phys. Commun. 
{\bf 105}, 42 (1997).
\bibitem{gs96} T.~Gehrmann and W.J.~Stirling, Phys. Rev. D {\bf 53}, 
6100 (1996).
\bibitem{jtst} T.~Sj\"ostrand {\it et al.}, Comput. Phys. Commun. 
{\bf 135}, 238     (2001). 
\bibitem{naomi} N.C.R.~Makins, HERMES, private communication (2002).
\bibitem{phd_frag} P.~Geiger, Ph.D. Thesis, University of Heidelberg (1998).
\bibitem{kk_phd} K.~Kowalik, Ph.D. Thesis, Institute for Nuclear Studies, 
Warsaw (2004).
\bibitem{hubert} H.~Gilly, Ph.D. Thesis, University of Freiburg (2000).
\bibitem{acta_paper} K.~Kowalik {\it et al.}, 
Acta Physica Polonica B {\bf 32}, 2929 (2001).

\end{thebibliography}
\end{document}